\pdfoutput=1
\documentclass{ws-jai}
\usepackage[flushleft]{threeparttable}
\usepackage{epstopdf}
\usepackage{hyperref}
\usepackage{float}

\begin{document}
\markboth{Smithsonian Astrophysical Observatory DSP Group}{SWARM: A 32~GHz Correlator and VLBI Beamformer for the SMA}

\title{SWARM: A 32~GHz Correlator and VLBI Beamformer for the Submillimeter Array}

\author{Rurik A. Primiani$^\dagger$, Kenneth H. Young$^\dagger$, Andr\'e Young$^\dagger$, Nimesh Patel$^\dagger$, Robert W. Wilson$^\dagger$, Laura Vertatschitsch$^\S$, Billie B. Chitwood$^\flat$, Ranjani Srinivasan$^\P$, David MacMahon$^\ddagger$ and Jonathan Weintroub$^\dagger$}

\address{
$^\dagger$Harvard-Smithsonian Center for Astrophysics, Cambridge, MA 02138, USA, rprimiani@cfa.harvard.edu\\
$^\ddagger$University of California Berkeley, Berkeley, CA 94720, USA\\
$^\S$Systems \& Technology Research, Woburn, MA 01801, USA \\
$^\flat$Smithsonian Astrophysical Observatory, Submillimeter Array, Hilo, HI 96720, USA \\
$^\P$Academia Sinica Institute of Astronomy and Astrophysics, Submillimeter Array, Hilo, HI 96720, USA \\
}
\maketitle

\catchline{}{}{}{}{} 


\begin{history}
\received{(to be inserted by publisher)};
\revised{(to be inserted by publisher)};
\accepted{(to be inserted by publisher)};
\end{history}

\begin{abstract}
A 32~GHz bandwidth VLBI capable correlator and phased array has been designed and deployed\footnote{At the time of writing three of four identical SWARM quadrants have been deployed supporting 24~GHz bandwidth. The full four-quadrant 32~GHz bandwidth system is expected to be completed by December 2016} at the Smithsonian Astrophysical Observatory's Submillimeter Array (SMA).  The SMA Wideband Astronomical ROACH2 Machine (SWARM) integrates two instruments: a correlator with 140~kHz spectral resolution across its full 32~GHz band, used for connected interferometric observations, and a phased array summer used when the SMA participates as a station in the Event Horizon Telescope (EHT) Very Long Baseline Interferometry (VLBI) array. For each SWARM quadrant, Reconfigurable Open Architecture Computing Hardware (ROACH2) units shared under open source from the Collaboration for Astronomy Signal Processing and Electronics Research (CASPER) are equipped with a pair of ultra-fast Analog-to-Digital Converters (ADCs), a Field Programmable Gate Array (FPGA) processor, and eight 10 Gigabit Ethernet ports.  A VLBI data recorder interface designated the SWARM Digital Back End, or SDBE, is implemented with a ninth ROACH2 per quadrant, feeding four Mark6 VLBI recorders with an aggregate recording rate of 64~Gbps. This paper describes the design and implementation of SWARM, as well as its deployment at SMA with reference to verification and science data.
\end{abstract}

\section{Introduction} \label{introduction}
The Submillimeter Array (SMA) is an eight-element radio interferometer located atop Mauna Kea in Hawai'i \cite{smaho}. Eight six-meter dishes may be arranged into configurations with baselines as long as 509~m, producing a synthesized beam of sub-arcsecond width at 345~GHz.

The SMA is expanding the bandwidth of its receiver sets to 8~GHz in each sideband. Two receivers can be operated simultaneously. Based on nominal center frequencies, the receivers are designated as 200, 240, 345, and 400.  The 200 and 240 are in opposite polarizations and can be tuned to overlap or to different bands, the same applies to the 345 and 400s.

To support the upgraded receivers a new wideband high spectral resolution correlator was needed. Scientific requirements called for 8~GHz bandwidth per sideband per polarization, 32 GHz bandwidth total, commensurate with the SMA's new receiver sets. High uniform spectral resolution of $\sim140~$kHz or finer over the entire band was specified to support fast spectral line surveys. Additionally, a phased array and VLBI data recorder were required to support Event Horizon Telescope (EHT) observations \cite{Johnson_2015}. The full set of scientific requirements is shown in Table \ref{tab:spec}. 

\begin{wstable}[htp]
\caption{SWARM's top level scientific requirements.}
\begin{tabular}{@{}lll@{}} \toprule
Feature  & Specification  & Remarks \\ \colrule
Number of antennas & 8 & dual frequency or dual polarization\\
IF bandwidth per quadrant & 4 GHz & 2 GHz per pol. per sideband\\
Total sky bandwidth & 32 GHz & 8 GHz per side band per pol\\
Simultaneous receivers & 2 &  dual freq. or dual pol. 230 \& 345 GHz  \\
Correlations & 128 & Full Stokes, $28\times4=112$ cross, $8\times2=16$ auto \\
Finest uniform resolution & 140~kHz  &  2.3 GHz Nyquist / 16384 channels \\
Fastest dump rate & 0.65 s & single full Walsh cycle\\
Phased array bandwidth & 16 GHz or 64 Gbps  & 4 GHz per sideband per polarization \\ \botrule
\end{tabular}
\label{tab:spec}
\end{wstable}

To meet the required specifications the SMA Wideband Astronomical ROACH2 Machine (SWARM) was envisioned, designed, and deployed. One quadrant of SWARM has 16 inputs, two receivers per SMA antenna, and can be configured to produce full Stokes polarization data over a 2 GHz usable band or a single Stokes polarization over a 4 GHz usable band. Thus, as a dual-sideband system four quadrants of SWARM will provide a total of 32~GHz of bandwidth on the sky. The system takes advantage of open source technology shared by the Collaboration for Astronomy Signal Processing and Electronics Research (CASPER)\footnote{For more information on CASPER, visit \url{http://casper.berkeley.edu}} as well as a five Giga-Sample-per-second (GSps) Analog-to-Digital Converter (ADC) board designed by the Academia Sinica Institute of Astronomy and Astrophysics (ASIAA) \cite{jiangadc}.

The Digital Signal Processing (DSP) platform chosen for SWARM is the second generation Reconfigurable Open Architecture Computing Hardware (ROACH2). Each of two channels per ROACH2 sample a baseband IF from a custom Block Down-Converter (BDC). The ADCs are clocked at 4.576~GSps thus sampling a 2.288~GHz Nyquist band corresponding to 2.000~GHz usable IF bandwidth per input (with excised guard-band). Each ROACH2 is host to one Xilinx Virtex-6 Field Programmable Gate Array (FPGA) chip which, when configured with the SWARM gateware, is host to two $32768$ point channelizers and a variety of other functions including fringe tracking, de-Walshing, full Stokes correlator, beamformer and packetized communication logic (for a detailed description of the gateware see Section \ref{fpga:intro}).


Although this paper is primarily about the new SMA correlator, SWARM, it is important to consider the context of the upgrade and the improvements that SWARM provides. SWARM will replace the recently-coined ASIC (Application Specific Integrated Circuit) correlator, previously called simply the ``correlator'', which unsurprisingly was built out of ASICs. An abridged list of SWARM benefits over the ASIC follows:
\begin{itemlist}
\item Higher uniform spectral resolution 
\item No trade-off between bandwidth and high spectral resolution
\item Large 2~GHz usable blocks are easier to passband calibrate, and result in superior spectra
\item Built in VLBI phased array processor and data storage, with 16$\times$ the present VLBI bandwidth
\item Better SNR due to more processed bits ($\sim 12\%$ in principal)
\item Smaller size and lower power consumption
\item Use of commodity components
\end{itemlist}
As the third and fourth quadrants of SWARM are installed and commissioned more of the ASIC correlator is removed. Eventually, by the end of 2016, only SWARM will remain. This phased upgrade from the old to the new correlator is intentional and permits a smooth transition for the SMA which must remain constantly in operation as an active facility instrument.
\section{System design}
The enormous computational requirements of SWARM demand a highly parallel signal processing engine. We selected the FPGA as the most appropriate technology. In particular CASPER hardware and libraries, along with the FX correlator architecture, have been qualified as the most viable design model.

CASPER's focus is on processing baselines for the very large numbers of stations common in modern low frequency radio arrays. The eight-antenna SMA is a modest size, but the extremely wide bandwidth presented an unexplored space within CASPER, and presents particular challenges. Early in the SWARM project we analyzed the resource requirements for the channelizers, which dominate the computational expense in the wideband FX architecture, and determined that the ROACH2 and in particular the Xilinx Virtex-6 SX475T FPGA would accommodate the SWARM gateware (see Section \ref{resource:estimate}).
\begin{figure}[tbh]
\begin{center}
\includegraphics[width=1.0\textwidth]{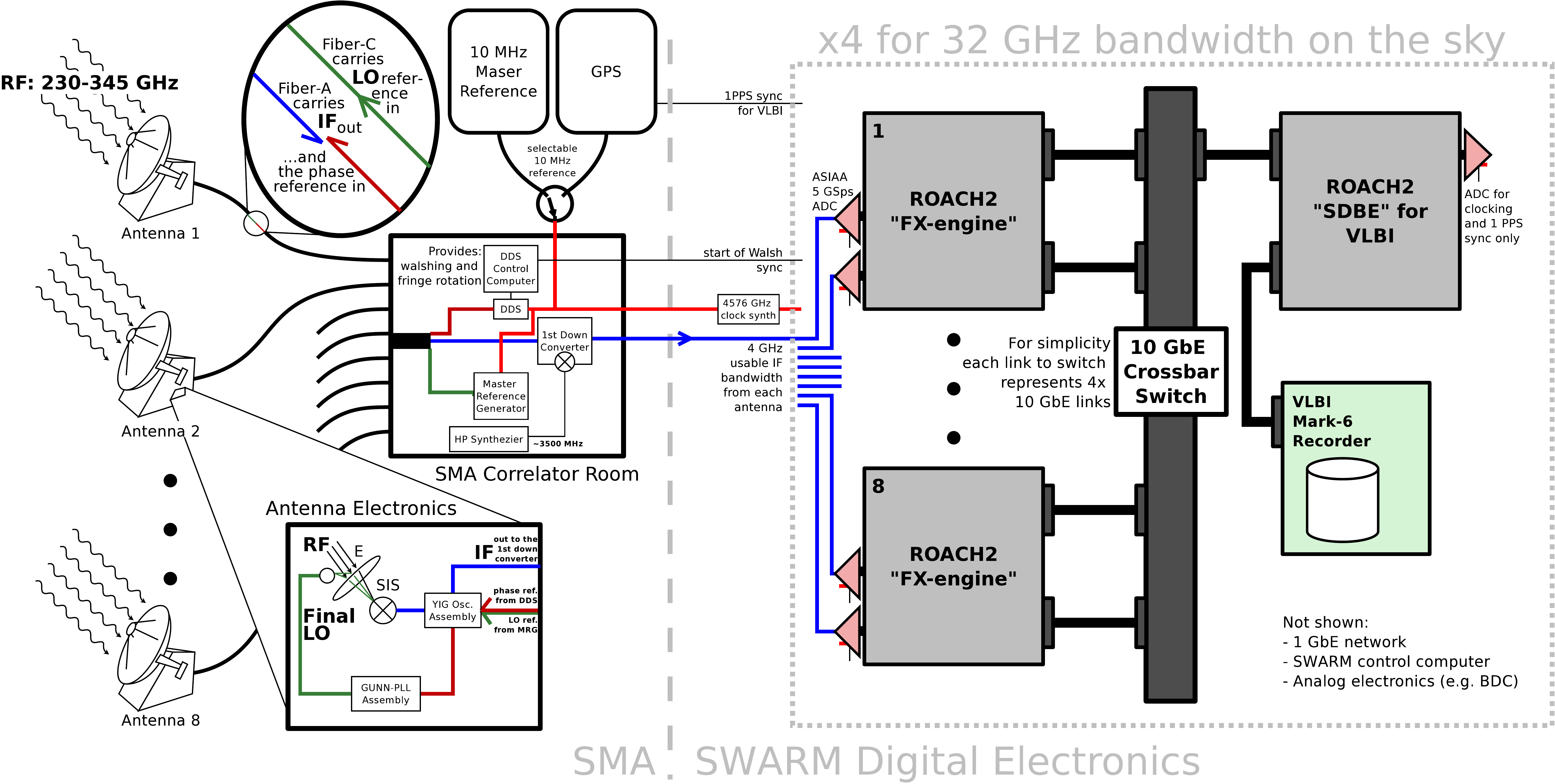}\\
\caption{Block diagram showing at the top level a quadrant of SWARM, on the right of the dotted line,  in the context of legacy SMA systems on the left.  There are eight ROACH2s on the left of the 10 GbE crossbar switch, which contain F- and X-engines, as well as coarse and fine delay tracking, phase control and deWalshing, a phased array summer, visibility accumulator, network logic, and assorted transposes and other memory.  On the right hand side of the switch is shown the ``SDBE'' and Mark6 data recorder, both required for EHT VLBI.}
\label{fig:swarmbd}
\end{center}
\end{figure}

\subsection{The CASPER packetized correlator} \label{sys:caspercorr}
CASPER pioneered the use of a commercial Ethernet switch as the interconnection fabric \cite{parsonsa}. Data is packetized prior to transmission via Ethernet switch ``crossbar'' from F-engine to X-engine and to VLBI recorders, etc.  

See Figure \ref{fig:swarmbd} which shows the architecture of a single SWARM quadrant at the top level, with the right hand side of the drawing showing the basic CASPER concept of processing engines organized around a 10 Gigabit Ethernet (GbE) switch. The usual CASPER architecture shows F-engines and X-engines (described in Sections \ref{fpga:f-engine} and \ref{fpga:x-engine}, respectively) on opposite sides of the switch, but in SWARM the F- and X-engines are folded back on one another, reducing the required number of ROACH2s by roughly two with almost the same reduction of required number of switch ports.

\subsection{Digital sampling}
As previously stated, we selected a CASPER compatible 5~GSps 8-bit ADC developed by our SMA partner, ASIAA, \cite{jiangadc} to process data in 2~GHz usable bandwidth blocks. The ASIAA ADC uses an integrated circuit ADC, the EV8AQ160, from {\it e2v}. This is a so-called quad core device, using four 1.25~GSps ADC cores interleaved to achieve the 5~GSps design rate. The device provides register controls to align the cores to reduce the impact of spurs which arise due to mis-alignment in offset, gain, phase (OGP), or threshold Integral Non-Linearity (INL). Top level specifications of the {\it e2v} are listed here (from EV8AQ160 data sheet\footnote{Available here: \url{http://www.e2v.com/resources/account/download-datasheet/2291}}):
\begin{itemlist}
\item Quad ADC with 8-bit resolution.
\item 5 GSps sampling rate in one-channel mode with four ADCs interleaved.
\item Digital interface (SPI) to set OGP and INL for individual cores.
\item Full power input bandwidth up to 2 GHz.
\item 500 mV peak-to-peak analog input.
\item SNR=44dB, ENOB=7.1 bit at 620~MHz input frequency
\end{itemlist}
In selecting the ADC, we required 2~GHz of usable bandwidth to support SWARM. A Nyquist zone up to 2.3 GHz is needed, with the upper edge of the usable 2~GHz band at 2.15~GHz. While the bandwidth of the {\it e2v} ADC in the data sheet is 2.0~GHz, our frequency response measurements show that the device responds beyond that limit, with the attenuation at 2.15~GHz about 6~dB (including any loss on the PC board). A sample rate of 4.6~GSps is thus within the maximum  specified of 5~GSps.

\subsubsection{Quad core calibration}
\begin{figure}[tbh]
\begin{center}
\includegraphics[width=0.7\textwidth]{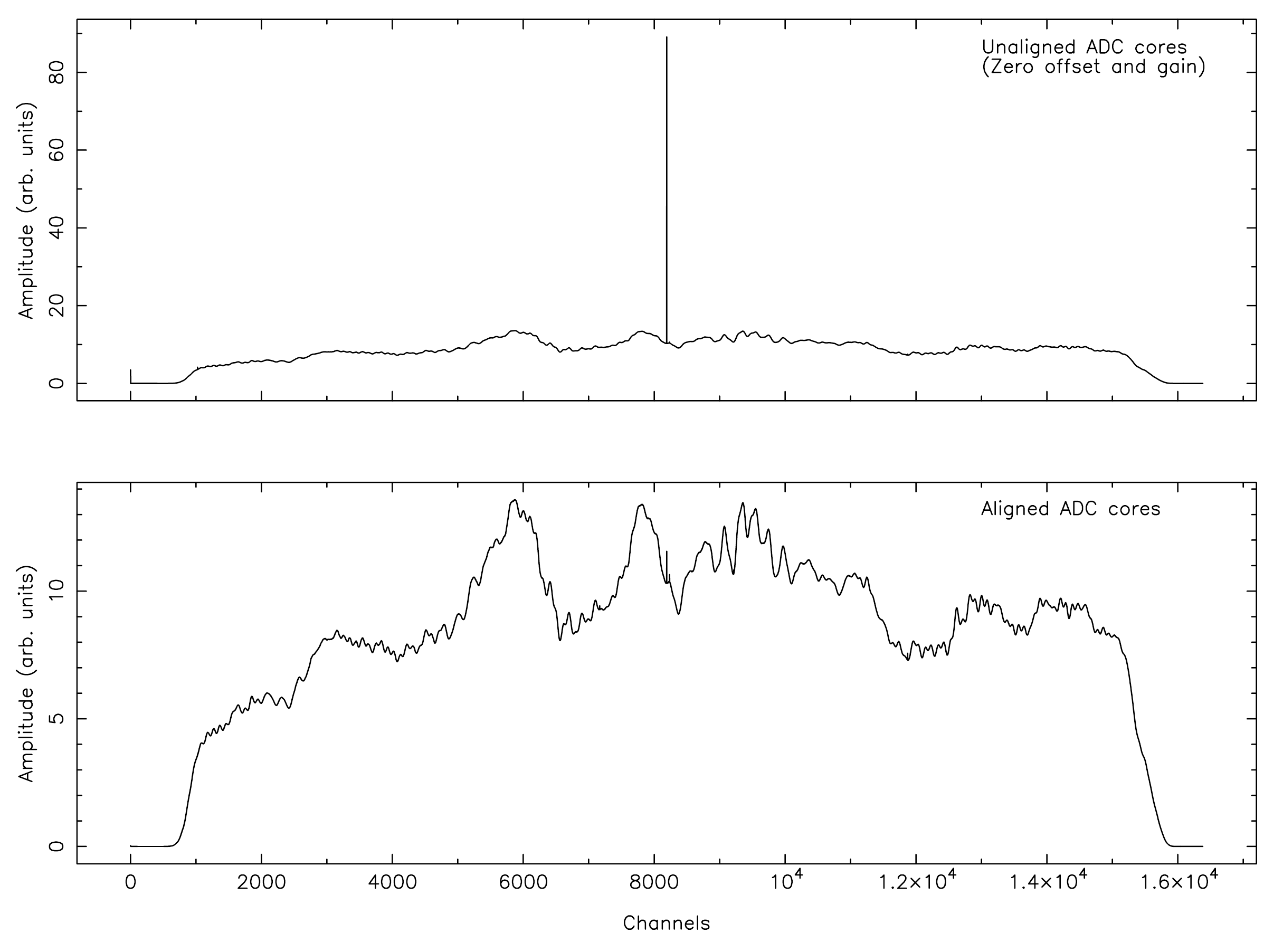}\\
\caption{Autocorrelation spectra obtained from one of the ADCs, over a 30 second integration, with the ambient temperature calibration load inserted. The top panel shows the spectrum with the offset and gain parameters set to zeroes, for the four cores of the ADC. It shows  a strong spur near the center, and a weaker spur  in the first channel. Setting the offset and gain values obtained from core alignment calibration with a noise source, removes the spurs effectively.}\label{fig:adcspur}
\end{center}
\end{figure}

\citet{patel} presented a series of measurements characterizing the performance of the ASIAA ADC. Signal-to-Noise and Distortion (SINAD), Spurious Free Dynamic Range (SFDR), Noise Power Ratio (NPR) and two-tone inter-modulation distortion tests showed that this ADC meets the requirements for SWARM. \citet{patel} also documents the quad core calibration methods used in characterizing the ADC using a sine wave source. One conclusion of our characterization of the ADC, however, was that the only core alignments that are critical for SWARM are offset and gain. When SWARM is installed at the SMA, the only input available without manual intervention is receiver noise. We have found that adjusting the offset and gain of the four cores to be equal using receiver noise provides adequate correction for our needs. Fig.\ref{fig:adcspur} shows the autocorrelation spectra obtained with one of the ADCs with the ambient temperature calibration load inserted at the 230 GHz receiver. With the offset and gain values set to zeroes for all four cores of the ADC, a strong spur is seen near the center of the spectrum. Drifts of the cores's offsets and gains have been small and slow.

\subsubsection{ADC power level} \label{sys:adcpow}
The optimal drive level into the ADC is determined by the peak of the Noise Power Ratio (NPR) curve vs. the input power. The NPR for an ideal 8 bit ADC (with only quantization noise and clipping noise) is 40.6 dB. The empirically determined NPR curve for the 5~GSps ADC boards used in SWARM can be seen in \citet{patel}. Patel measures the NPR curve using a tunable notch filter set to frequencies of 800~MHz, 1000~MHz, and 1750~MHz, all of which show good agreement with the theoretical curve but with peak NPR degrading slowly with frequency, $\sim$3.6~dB/GHz. While a Loading Factor (LF) of -11~dB yields the highest possible NPR, corresponding to the peak at 800~MHz, the degradation at the higher frequencies was deemed unacceptable. Instead the peak of the 1750~MHz measurement was chosen in order to  optimize both NPR and NPR-slope with frequency; this corresponds to a LF of -12.5~dB, or equivalently a drive level of -14.5~dBm.

A two-stage software servo was developed for the BDC to maintain proper power levels. The initial closed loop servo ensures that, firstly, the input IF power does not compress the mixer stage that converts IF to baseband and, secondly, that the output baseband stage attenuators set the power level going into the ADC to roughly the LF that was determined to be optimal. The post-servo, open-loop leveling that executes continuously keeps the LF at -12.5 dB. The correction occurs once per second. This ensures that the drive levels into the ADC stay constant throughout the night and are impervious to system temperature changes. A peak-to-peak fluctuation of 0.3-0.4~dB from the nominal value is considered satisfactory. 
 
\subsection{ROACH2} \label{sys:roach2}
The latest open-source DSP platform to come out of CASPER is the so-called ROACH2. It is built into a 1U ATX computer format and hosts a Xilinx Virtex 6 SX475T FPGA and a PowerPC.\footnote{For a detailed block diagram of the ROACH2 platform, visit \url{http://casper.berkeley.edu/wiki/ROACH2}} ROACH2 has two expansion connectors that are typically used to connect ADC cards. It uses the FPGA for its processing element and has additional memory for storage and a PowerPC unit for monitor and control. It also has 80~Gbps of bidirectional digital interface bandwidth.

\subsection{High speed network crossbar}
The CASPER correlator architecture uses processing nodes that communicate via packetized data routed through commercially available switches. These nodes could be FPGA, ASIC, GPU or CPU/multicore based depending on the specific requirements for the node and the maturity of the instrument. 
\begin{figure}[tbh]
\begin{center}
\includegraphics[width=0.7\textwidth]{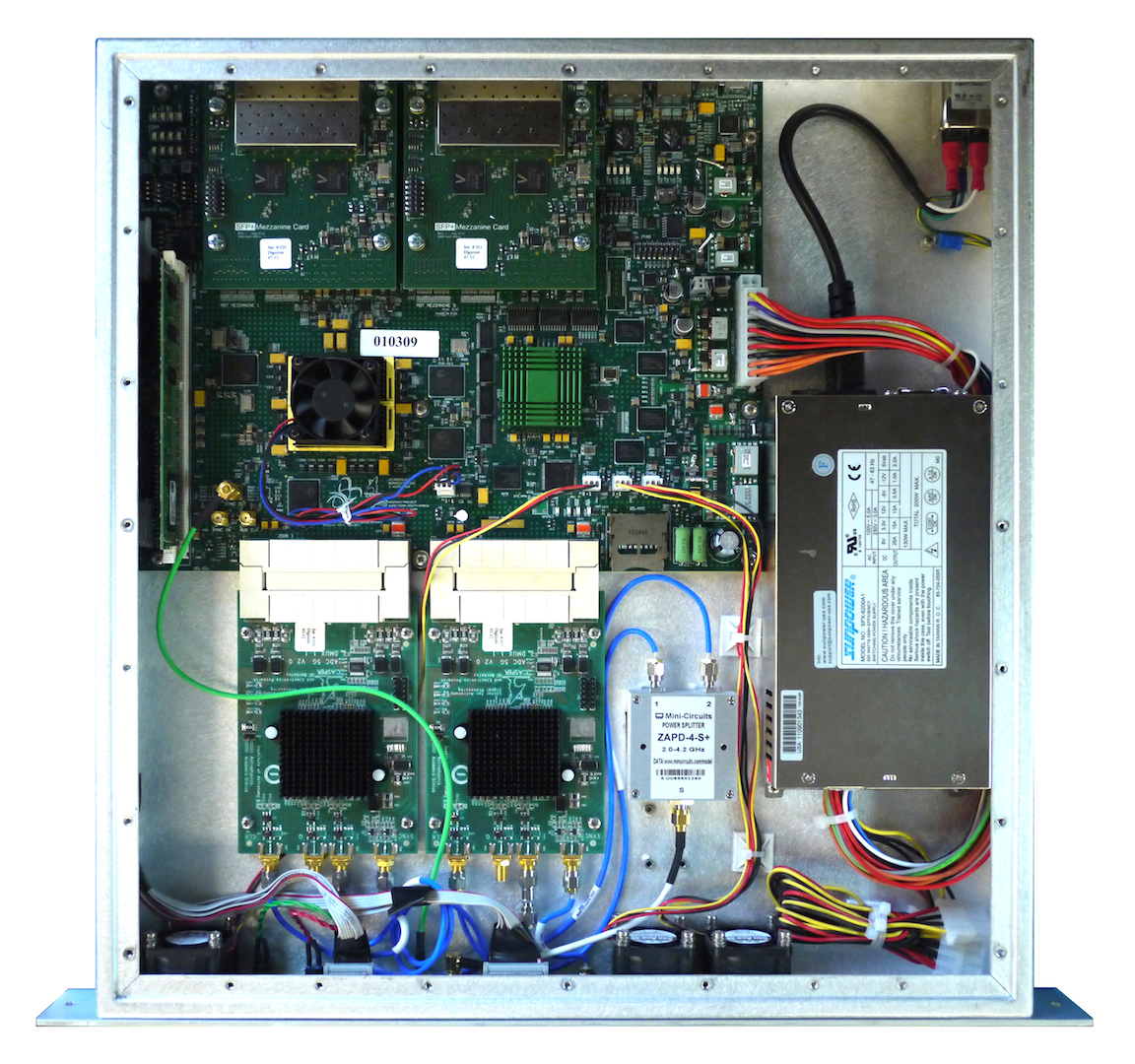}\\
\caption{Plan view photo of the ROACH2 platform configured for SWARM.  Two 5~GSps Quad Core ADCs are plugged in to the ZDOK connectors towards the bottom, providing samples at a data rate approaching 80~Gbps.  Eight 10 GbE ports on the mezzanine board towards the top provide matched data rate throughput to the network switch. Photo credit: Derek Kubo.}
\label{fig:pheff}
\end{center}
\end{figure}

\subsection{Cooling the electronics}
Operation of the ROACH2 chassis at the SMA facility near the summit of Mauna Kea, at an elevation of approximately 4000 meters or 13,000+ feet, presented difficulties not present at sea level laboratory testing. In particular, the FPGA die temperature quickly exceeded the 85 degrees C threshold for guaranteed timing, even at reduced clock speeds. Modifications were made to the ROACH2 chassis to divert the power supply exhaust and to increase air flow through chassis from front to back. The cooling fan for the FPGA heat sink was increased in power and changed in orientation, and a more effective heatsink compound was utilized. 

Modifications were also made to the 19 inch equipment racks that house SWARM hardware. The previously open racks were enclosed with front and rear doors, and refrigerated air passing from the bottom of the rack was deflected into an added front plenum, whence the cooled air was then drawn, through the front of SWARM components, including the ROACH2 chassis, and exhausted out the rear, to be carried up and out of a damper controlled exhaust at the top. The combination of these modifications allowed the FPGA die temperature to remain below 85 degrees C at the full clock rate. 

\subsection{Real-time software}
Although often overlooked and underestimated, real-time software is critical to smooth operation of an array. For SWARM much of the pre-existing SMA software environment was adapted for the monitor and control of the new correlator. This included reuse and modification of:
\begin{itemlist}
\item Direct Digital Synthesizer (DDS) control code that manages Walshing and fringe-rotation
\item Shared-memory library for sharing values between SWARM and the SMA software environment
\item Correlation plotter for displaying SWARM data alongside the ASIC correlator data.
\item Data archive software for storing SWARM data using the existing SMA data format.
\end{itemlist}
Additionally, some new software was developed in Python\footnote{For more on the Python programming language, visit \url{http://www.python.org}} for receiving and reordering of the SWARM visibility dumps as well as for VLBI phased-array calibration (see section \ref{vlbi:beamformer}). As previously discussed in Section \ref{sys:adcpow} the software servo for the BDC was also written to run in real-time.

\section{FPGA gateware} \label{fpga:intro}
Each ROACH2 board in SWARM contains a single FPGA connected to multiple peripherals (for more information on ROACH2 see Section \ref{sys:roach2}). The FPGA logic is implemented using what is called a \emph{bitcode} which is essentially a binary file that encodes the configuration of logic elements on the chip and the connections between them. Typically the bitcode is generated by starting with a high-level description of the intended behavior; this is then synthesized and mapped to the logic elements provided by the FPGA. For the purposes of this document we will refer to this high-level description as \emph{gateware}.

\begin{figure}[tbh]
\begin{center}
\includegraphics[width=1.0\textwidth, trim=0 150px 0 0, clip]{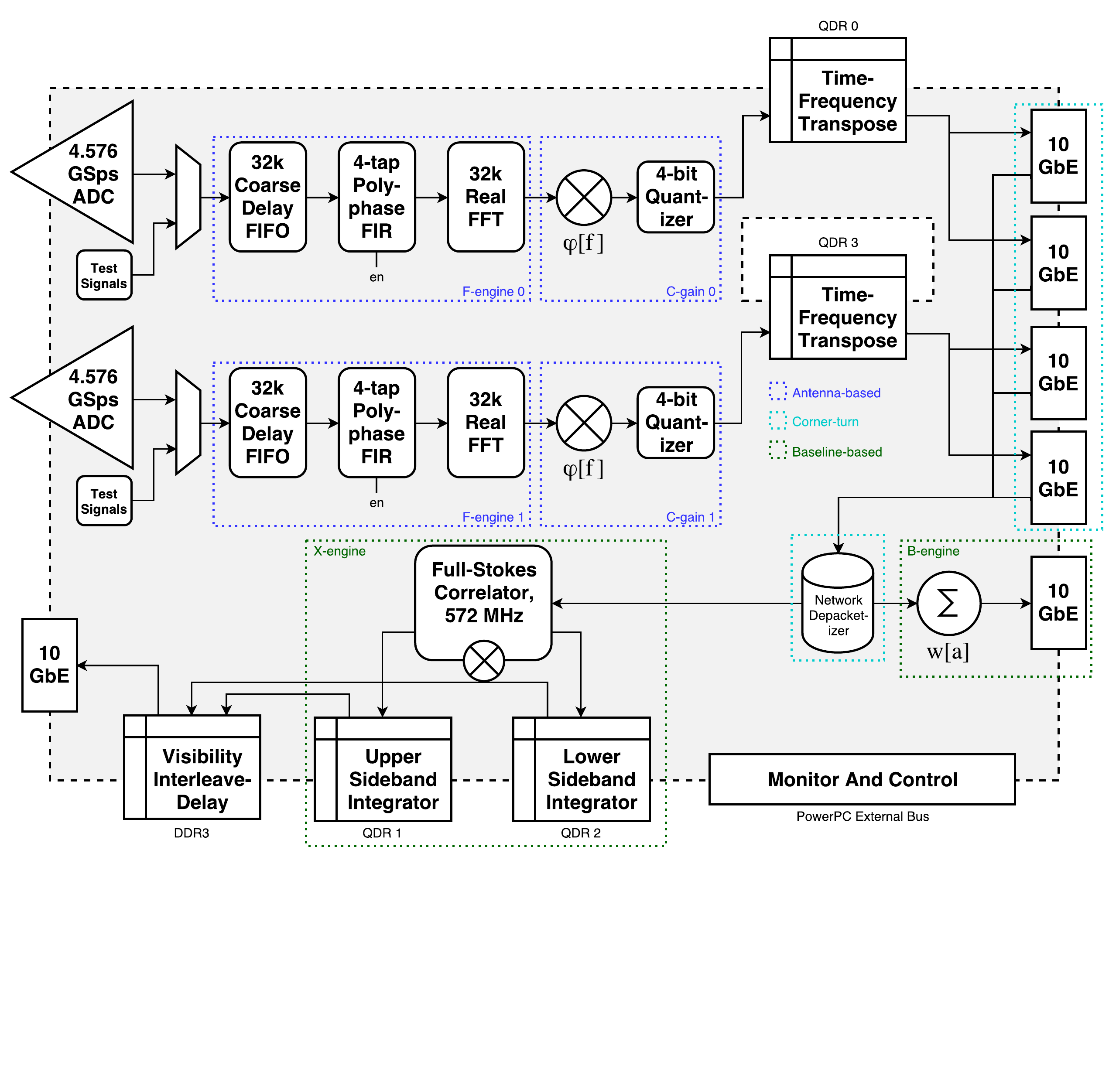}\\
\caption{Block diagram representing the SWARM gateware design. The dashed, shaded region represents the FPGA on the ROACH2 platform, a Virtex-6 SX475T. Blocks fully inside this shaded region represent high-level logic within the gateware while blocks bordering it represent external interfaces (e.g. memory controllers, network ports, and busses). Dotted regions identify sub-systems referred to throughout this document with their hierarchical name. The data from each antenna flows from the ADCs on the left through the two F-engines, gets time-frequency transposed by two Quad Data Rate (QDR) chips, is sent out over the network to return as $1/8$ the bandwidth but for all $8$ antennas, is then correlated by the X-engines, integrated using a QDR per sideband, and finally sent out again over the network to be archived.}
\label{fig:swarmgw}
\end{center}
\end{figure}

Although it is common (i.e. in the engineering industry) for gateware to be implemented using languages such as Verilog or VHDL, for SWARM we decided to take advantage of the large and open-source CASPER gateware library and toolflow based around the MATLAB Simulink\footnote{For more on MATLAB Simulink, visit \url{http://www.mathworks.com/products/simulink}} design environment. This decision had huge advantages including allowing us to significantly reduce development by designing at a very high-level. For example, the CASPER libraries provide parameterized blocks for a Fast Fourier Transform (FFT). What follows is a detailed description of the SWARM gateware\footnote{All source code for gateware and related software is hosted on Github at \url{http://www.github.com/sma-wideband}} which is graphically described in the block diagram in Figure \ref{fig:swarmgw}.
%

\subsection{Selectable test signals} \label{fpga:source}
The SWARM gateware design features a software-selectable data source which defaults to the 8-bit data from the samplers but can be selected to be either a Gaussian noise generator, a tunable sine-wave test tone, or a summation of both; the two inputs paths are independently selectable. In practice the Gaussian noise is used to verify basic functionality of the system from the inputs to the outputs by first selecting the noise for all inputs, then synchronizing them across all DSP boards, and verifying perfect correlation on all baselines of the visibility output data. This first-pass test has proven to be very helpful in quickly diagnosing issues throughout the design.

\subsection{F-engine and coarse-delay} \label{fpga:f-engine}
Fundamental to the FX correlator architecture is the conversion of a sequence of discrete time domain samples, for every input, to the Fourier domain before they are cross-correlated; this is referred to as the ``F-engine.'' In the SWARM FPGA gateware design there are two such F-engines instantiated which separately process either two contiguous frequency bands or two orthogonal polarizations per antenna depending on the mode. Each quadrant of SWARM thus has 16 F-engines for a total of 64 across all quadrants.

The SWARM F-engine is, in practice, a Polyphase Filter Bank (PFB) implemented with a $32768$-point real-valued FFT preceded by a 4-tap Hamming-window Finite Impulse Response (FIR) filter for each polyphase component. Although the PFB provides the best isolation for narrow spectral components the SWARM F-engine features the ability to disable the FIR at runtime for observations that would prefer a straight FFT (such as VLBI where easy conversion back to the time domain is necessary). Both the FIR and the FFT are implemented using standard blocks from the CASPER library with a parallelization factor, i.e. ``demux,'' of 16. Ultimately, the output of the F-engines are complex spectra of $16384$ channels for every $32768$ input time-domain samples; at a sample rate of 4576~MHz that amounts to a transformation roughly every 7 microseconds.

In order to align the F-engine windows the SWARM gateware includes a coarse-delay correction which is applied before the PFB using a buffer in the time domain. The primary purpose of the coarse-delay is to correct for the large geometric delays between antennas when tracking a celestial source. To accommodate the largest baselines of the SMA in the Very Extended (VEX) configuration the buffer is $32768$ samples, coincidentally equal to one FFT window.

\subsection{Fine-delay, phase, and amplitude control} \label{fpga:cgain}
Directly following the F-engines are the so-called ``complex gain'' blocks which multiply each channel of every spectra by a dynamic complex value. There is a single fine-delay control, a phase control, and a per-channel amplitude control. The fine delay control amounts to simply a phase-per-channel value while the phase control is a constant phase across the band (i.e. every channel gets the same phase). There is also a per-channel amplitude control implemented using a software-accessible memory bank. In practice the amplitude control is rarely used (since most amplitude bandpass variations can be calibrated using bandpass calibration sources) but could be used to optimize the secondary quantization (see Section \ref{fpga:quant}) as well as to knock out sources of interference (such as leakage from the oscillators in the antenna electronics).

\subsection{Synchronization and de-Walshing} \label{fpga:sync}
The SMA uses Walsh modulation and demodulation to reduce cross-talk within the IF/LO system as well as for sideband separation in the correlator. The modulation is applied at the LO while within SWARM Walsh demodulation is done using the complex gain adjustments discussed in the previous Section \ref{fpga:cgain}. The modulation and demodulation are synchronized using an external signal generated by the DDS-computer (the machine that handles the modulation of the LO). Within the SWARM gateware the external signal drives an arm-able internal Walsh counter which is then used to demodulate both the 0-180 degree (for cross-talk rejection) and the 90-270 degree (sideband separation) components of the Walsh pattern. The input signals are fully demodulated for one sideband, typically the USB, via phase shifts applied with the complex gain sub-system (see Section \ref{fpga:cgain}); subsequently the other sideband is ``separated'' in the final accumulator (see Section \ref{fpga:visibs}) by accumulating a parallel integration with a secondary modulation opposite to that of the USB (on a per-baseline basis).

\subsection{Quantization to 4-bits} \label{fpga:quant}
To reduce the memory and network-traffic requirements of the transpose and corner-turn operations (see Section \ref{fpga:transposes}) while maintaining signal-to-noise and dynamic range it was decided to re-quantize to lower resolution after the complex gains are applied. Although the samplers themselves provide 8-bits the data grows to 18-bits through the F-engine and the complex gain subsystems but is subsequently rounded down to 4-bits. This bitwidth is common among packetized CASPER-based FX correlators.

\subsection{Time-Frequency Transposes} \label{fpga:transposes}
The F-engines' output is a continuous series of spectra, however the X-engines, i.e. the correlators, expect a sequence of time samples for each channel (see Section \ref{fpga:x-engine}). To reorder the F-engine outputs appropriately SWARM uses the high-speed QDR memory provided by the ROACH2 boards. The X-engines expect 128 times samples per channel, thus 128 spectra from each F-engine must be buffered row-wise while the per-channel data is read out column-wise. This process is effectively a matrix transpose operation where the axes being transposed are frequency and time. In practice the spectra are actually ``double-buffered'' in the QDR memory (for simplification of read/write addressing) thus requiring a total of 
\[
2 \times 128 \textrm{ spectra} \times 16384 \textrm{ channels} \times 2 \textrm{ components (real/imaginary)} \times 4 \textrm{ bits} \approx 4.2 \textrm{ MB}
\]
Although each QDR memory on the ROACH2 board has a roughly $9$ MB capacity SWARM uses one QDR for each F-engine path; this leaves plenty of headroom and leaves open the possibility of doubling the F-engine channels in the future (halving the continuum resolution).

\subsection{Packetized Corner-turn} \label{fpga:cornerturn}
Once the frequency domain data has been time-frequency transposed it must be transposed in another way, frequency-antenna. On one side, each F-engine path produces a full spectra for a single antenna while on the opposite end a single X-engine will consume some subset of the channels (i.e. bandwidth) for \emph{all} antennas. This process is commonly referred to as the ``corner-turn'' and is a requirement for any correlator.

A corner-turn can be implemented in numerous ways. The ALMA correlator, for example, uses $16384$ cables to route the data appropriately which turned out to represent the ``greatest design challenge in the system'' \cite{escoffier}. SWARM, on the other hand, uses what could be called the ``CASPER approach'' (see Section \ref{sys:caspercorr} and Figure \ref{fig:swarmbd}) which is to use a commercial high-speed Ethernet switch and routed packets to serve the same function. Although some overhead is needed in the gateware to accommodate packet buffers this approach has the benefit of being flexible, highly-scalable, easier to implement, and, in many cases, can be cheaper than other methods.

\subsection{X-engine and accumulators} \label{fpga:x-engine}
Once the data has been corner-turned each processor now has access to data from all antennas for a subset of the bandwidth which for SWARM is $1/8$ of the channels. This means that all baselines can be formed and baseline-based processing can begin. In particular all inputs can now be correlated by a subsystem called the X-engine. For the SWARM gateware we use the standard CASPER library X-engine block, eight per board due to the 16-fold demux (the factor of two comes from having complex-valued data after the F-engine).

Unlike in many other CASPER correlators the SWARM X-engines are co-located with the F-engines, that is to say they use the same processing boards as the F-engines. While this has presented challenges in terms of clocking the FPGA design at high clock rates the approach was intended to reduce the total number of ROACH2 boards (thus reducing cost) as well as using fewer Ethernet switch ports for the corner-turn. Additionally the corner-turn switch ports are all used full-duplex at very nearly 10~Gbps in both directions.

The SWARM X-engines compute all cross-correlation products regardless of whether the two inputs per SWARM board represent two polarizations, i.e. dual-polarization mode, or two contiguous chunks of bandwidth, i.e. single-polarization mode. So, although we consider SWARM to be an 8-element full-Stokes correlator it could also be thought of as a 16-element single-Stokes correlator. Additionally the auto-correlations are produced which have proven useful for calibrating the data. In total, the X-engines produce 120 complex-valued cross-correlations and 16 real-valued auto-correlations; each pair of real-valued auto-correlations can be crammed into a single complex number thus reducing the total output components to 128.

For efficient use of resources the X-engine blocks are configured to integrate by 128 time samples and the outputs from all eight X-engine blocks (which simultaneously compute eight different channels) are interleaved into a single stream. This data is then long-term accumulated using one QDR chip per sideband as discussed in Section~\ref{fpga:sync}.

Note that the input window for each X-engine block is 1024 clocks (128 samples for each antenna-receiver pair) while the valid output window is 128 clocks (one clock per component). However because we're interleaving eight blocks going into the accumulator there is no idle time available for double-buffering (though the capacity \emph{is} available) and therefore the accumulations must be read out immediately upon completion. This presents a particular challenge for reading the data across an entire SWARM quadrant, the solution for which is discussed in Section~\ref{fpga:visibs}.

\subsection{Visibility output and interleave delay} \label{fpga:visibs}
All ROACH2s are synchronized using an external signal, this applies to the X-engines as well. Thus, the X-engines all dump their visibility data simultaneously. While the average data rate at this point in the system is small (typical integration times are $\sim$30 seconds), the simultaneous transmission of this data to a single port connected to a control computer would overwhelm the limited internal memory buffer in SWARM's 10 GbE switch. The solution was to add a software-defined delay to the FPGA gateware in order to stagger the X-engine outputs. Due to the large size of the visibility data, this required using the on-board DDR3 memory, which offers 4 GB of memory. 

\subsection{B-engine} \label{fpga:b-engine}
Another baseline-based system is the built-in beamformer that enables the SMA to operate in a phased array mode, called the B-engine. The beamformer provides an adjustable gain per antenna which can be used effectively as a mask and sums all antennas (a reduction of the data rate by eight). The summed data is then sent out onto the network to the VLBI processor and recorder. To effectively be used as a phased-array for VLBI the phases need to be adjusted for each antenna in real-time using the constant phase component of the complex gain subsystem (see Section \ref{vlbi:beamformer}).

\section{Resource utilization} \label{resource:intro}
Before committing to the ROACH2 we needed to understand that the bitcode would fit the target FPGA. It was clear that the utilization would be dominated by the PFB. A hard reality is that the FPGA cannot run nearly as fast as the ADC so it is necessary to process a number of parallel streams, the demux factor. An early SWARM Memo\footnote{SWARM Memo \#1 \url{https://www.cfa.harvard.edu/twpub/SMAwideband/MemoSeries/sma_wideband_utilization_1.pdf}} explored the resource requirements of the PFB which was vital in the decision to proceed with using the ROACH2 for the SWARM project. This section will reproduce (but not derive) the results from that Memo.

\subsection{Estimation of resources} \label{resource:estimate}
As discussed in Section \ref{fpga:f-engine}, a PFB can be constructed using an FIR filter followed by a DFT which extracts the appropriate sub-bands. The DFT can be implemented using a FFT algorithm in order to take advantage of the $O(N\log N)$ optimization those algorithms afford. However as bandwidth, and therefore demux (represented here as $D$), grows, more samples are presented at once which means more multipliers must be instantiated in hardware.

\begin{figure}[htp]
\centering
\includegraphics[width=0.8\textwidth]{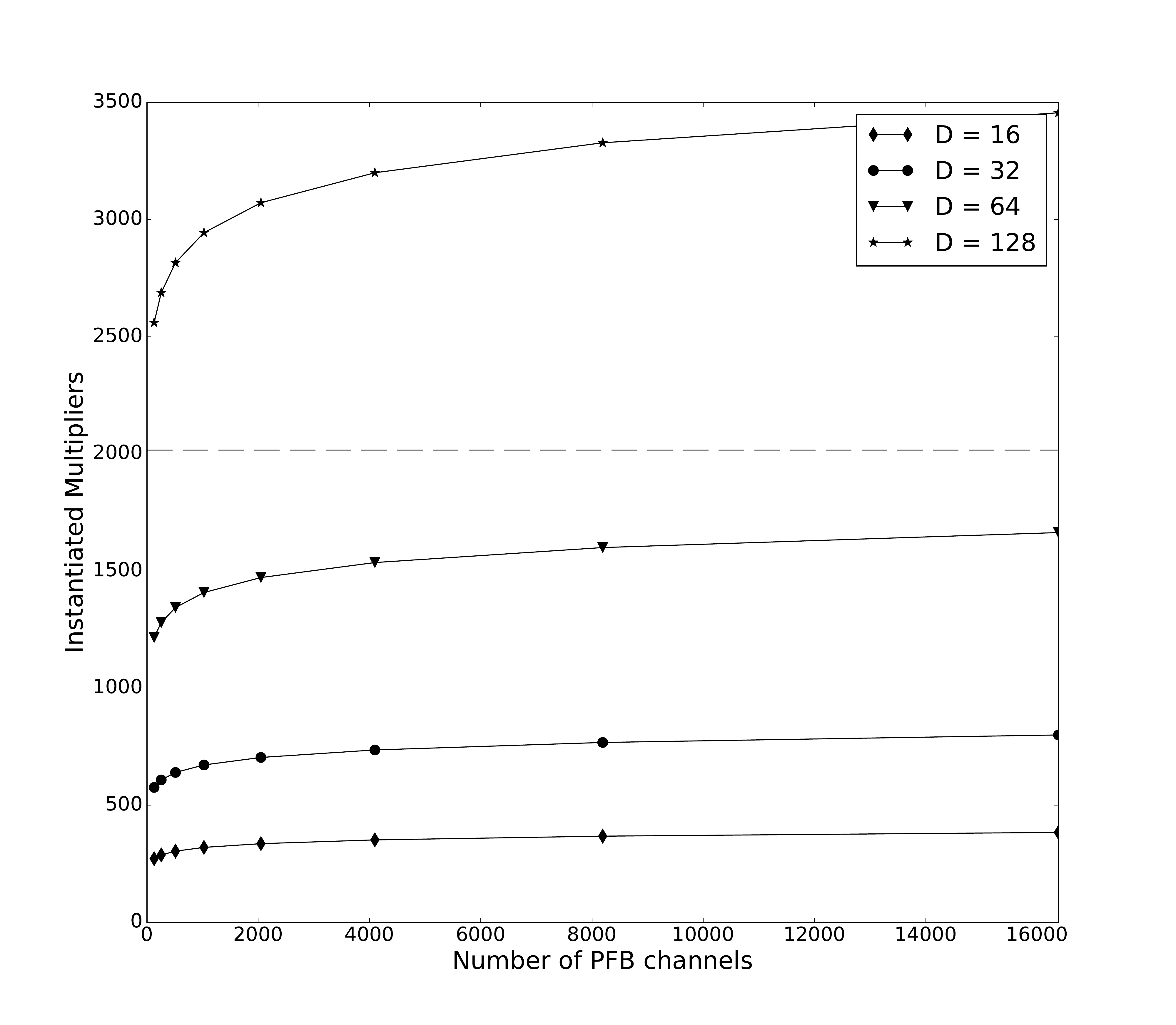}
\caption{Number of multipliers versus number of PFB channels for various values of the demultiplex factor. The dashed line represents the upper limit of available DSP slices on the FPGA present on the ROACH2. This plots shows how expensive it can be to jump to the next demux (e.g. to decrease the FPGA clock speed) while maintaining the same number of channels. Note: here we are assuming 8 taps in the FIR (where as SWARM uses only 4).}
\label{mults_vs_channels}
\end{figure}

The FIR filter preceding the FFT uses a single real multiplier per tap, so given $T$ taps (typical numbers are 4 to 8) and $N$ channels, the full PFB multiplier utilization is shown below with the various components identified,
\begin{equation}
\label{resource:mults}
M_\mathrm{PFB} = \underbrace{D\log_2 ND}_{\mathrm{FFT}} + \underbrace{TD}_{\mathrm{FIR}} - \underbrace{2D}_{\mathrm{optimization}}
\end{equation}
The most important thing to note from this equation is that since the pipelined stages divide their computation over $N/D$ clock cycles the total number of multipliers only grows with $N$ as $\log N$. The main contributer to multiplier utilization is instead the demux factor. Figure \ref{mults_vs_channels} shows a graphical representation of Equation~\ref{resource:mults}.
To find the total adders we go through a similar calculation but noting that the butterflies have $\frac{3}{2}$ as many adders as multipliers and the FIR only needs to sum all taps and thus performs $D(T-1)$ adds. The total is shown below, again with the contributions pointed out,
\begin{equation}
\label{aresource:adds}
A_\mathrm{PFB} = \underbrace{\frac{3}{2}D\log_2 ND}_{\mathrm{FFT}} + \underbrace{D(T-1)}_{FIR} + \underbrace{D}_{\mathrm{reorder}}
\end{equation}
The added $D$ results from extra adders in the block needing to do the final reordering of the FFT output. 

Within a polyphase filter-bank the multiplier and adder utilization appear to grow significantly with the demux factor, namely as $D\log_2{ND}$, whereas the amount of required memory depends critically, and linearly, on the total channels, $N$. Generally this implies that designs with modest bandwidth but requiring significant spectral resolution will be constrained by memory. SWARM however, has both very large bandwidth and substantial PFB size to achieve fine spectral resolution. This formalism helped us find the appropriate combination of parameters which meet the requirements of bandwidth and spectral resolution while fitting the logic and memory available in the ROACH2's FPGA.

\subsection{Demultiplexing} \label{resource:demux}
Experience shows that clocking an FPGA with a complex bitcode at rates approaching or exceeding 300~MHz stretches its capabilities, and those of the design tool-flow, to meet timing. Were 312~MHz achievable, however, our resource calculation shows that a very substantial savings in multiplier and adder resources results (see Figure~\ref{mults_vs_channels}). Without constraint it would perhaps be preferred to clock the FPGA at about 250~MHz, however because the demux factors are quantized to radix-2 numbers, it is important to appreciate that stretching to the next demux boundary can yield significant returns in utilization.

\subsection{Implementation resources used} \label{resource:used}
Ultimately the SWARM gateware described in Section \ref{fpga:intro} fit into the target FPGA with a demux factor of $16$ which meant clocking the FPGA at 286~MHz. For a full list of resources used by the implementation of our gateware, divided by sub-system, see Table \ref{fpga:resource-table} below.

\begin{wstable}[htp]
\caption{Resources used by various sub-systems of the SWARM gateware.}
\begin{tabular}{@{}llllll@{}} \toprule
                   & DSP Slices   & Slice LUTs      & Slice Reg.      & Block RAM    & Slices         \\ \colrule
Available\tnote{a} & 2016         & 297600          & 595200          & 1064         & 74400          \\ \colrule
F-engine 0     & 336 (16.7\%) & 66546  (22.4\%) & 74637  (12.5\%) & 232 (21.8\%) & 21059 (28.3\%) \\
F-engine 1     & 336 (16.7\%) & 66756  (22.4\%) & 74637  (12.5\%) & 232 (21.8\%) & 20700 (27.8\%) \\
Complex gain 0 & 64  (3.2\%)  & 3929   (1.3\%)  & 3640   (0.6\%)  & 64  (6.0\%)  & 1876  (2.5\%)  \\
Complex gain 1 & 64  (3.2\%)  & 3909   (1.3\%)  & 3641   (0.6\%)  & 64  (6.0\%)  & 1640  (2.2\%)  \\
X-engine       & 4   (0.2\%)  & 44475  (14.9\%) & 47244  (7.9\%)  & 95  (8.9\%)  & 10394 (14.0\%) \\
B-engine       & 42  (2.1\%)  & 1646   (0.6\%)  & 2101   (0.4\%)  & 10  (0.9\%)  & 868   (1.2\%)  \\
Other          & 64  (3.2\%)  & 55297  (18.6\%) & 58479  (9.8\%)  & 200 (18.8\%) & 16988 (22.8\%) \\ \colrule
Total          & 910 (45.1\%) & 242558 (81.5\%) & 264379 (44.4\%) & 897 (84.3\%) & 73525 (98.8\%) \\ \botrule
\end{tabular}
\begin{tablenotes}
\item[a] Shown for the Xilinx Virtex-6 SX475T.
\end{tablenotes}
\label{fpga:resource-table}
\end{wstable}

\section{VLBI features}
SWARM supports VLBI through a built-in beamformer, a VLBI-specific packetizer called the SWARM Digital Backend (SDBE), and an off-line data preprocessing system called the Adaptive Phased-array and Heterogeneous Interpolating Downsampler for SWARM (APHIDS). This enables the SMA to participate in VLBI observations as part of the EHT.

\subsection{Beamformer} \label{vlbi:beamformer}
The beamformer coherently adds the signals received from the target source in each antenna such that the array performs as the equivalent of a single station with a larger collecting area within the wider VLBI array. Phasing the array requires tracking all sources of delay, including fluctuations in water vapor concentration in the atmosphere. The SWARM phasing system is equipped with a real-time phasing solver that continually updates the beamforming weights to compensate for these variable delays, which manifest as variable phase errors in each antenna, over the course of the observation. Since the phased array capability is used to observe sources that are unresolved on baselines within the array, the corrective beamformer weights can be computed by extracting from the correlator output that contribution associated with a point-like source. Furthermore, as the weights are applied to the signal from each antenna before computing cross-correlations between antenna pairs (see Figure~\ref{fig:swarmgw}), the solution obtained from the correlator output for a particular integration period can also be used to calculate the average phasing efficiency over that same period. Specifically, the phasing efficiency is calculated as,
\begin{equation}
  \eta_\phi = \left| \sum_i w_i \right|^2 \bigg{/} \left( \sum_i \left| w_i \right| \right)^2,
\label{eq:pheff}
\end{equation}
where $w_i$ is the complex-valued weight applied to antenna $i$. See \citet{Young16} \footnote{See also the related SMA Memo 163 at \url{https://www.cfa.harvard.edu/sma/memos/163.pdf}} for a more detailed discussion of the phased array and performance assessment thereof.

Figure~\ref{fig:pheff} shows the phasing efficiency achieved over the course of several scans during one night of the 2016 EHT campaign. For most of the scans the efficiency is well above 0.9. Lower values obtained during the scans on Cen~A (just after 8:00~UT) and the first few scans on SgrA* and NRAO~530 (from 11:00~UT) are attributed to observing at low elevation which degrades the atmospheric phase stability. The antenna that was used as the phase reference during the observation suffered a loss of coherence from around 10:00--11:00~UT which resulted in poorer performance for scans in that period.

\begin{figure}[tbh]
\begin{center}
\includegraphics[width=0.8\textwidth]{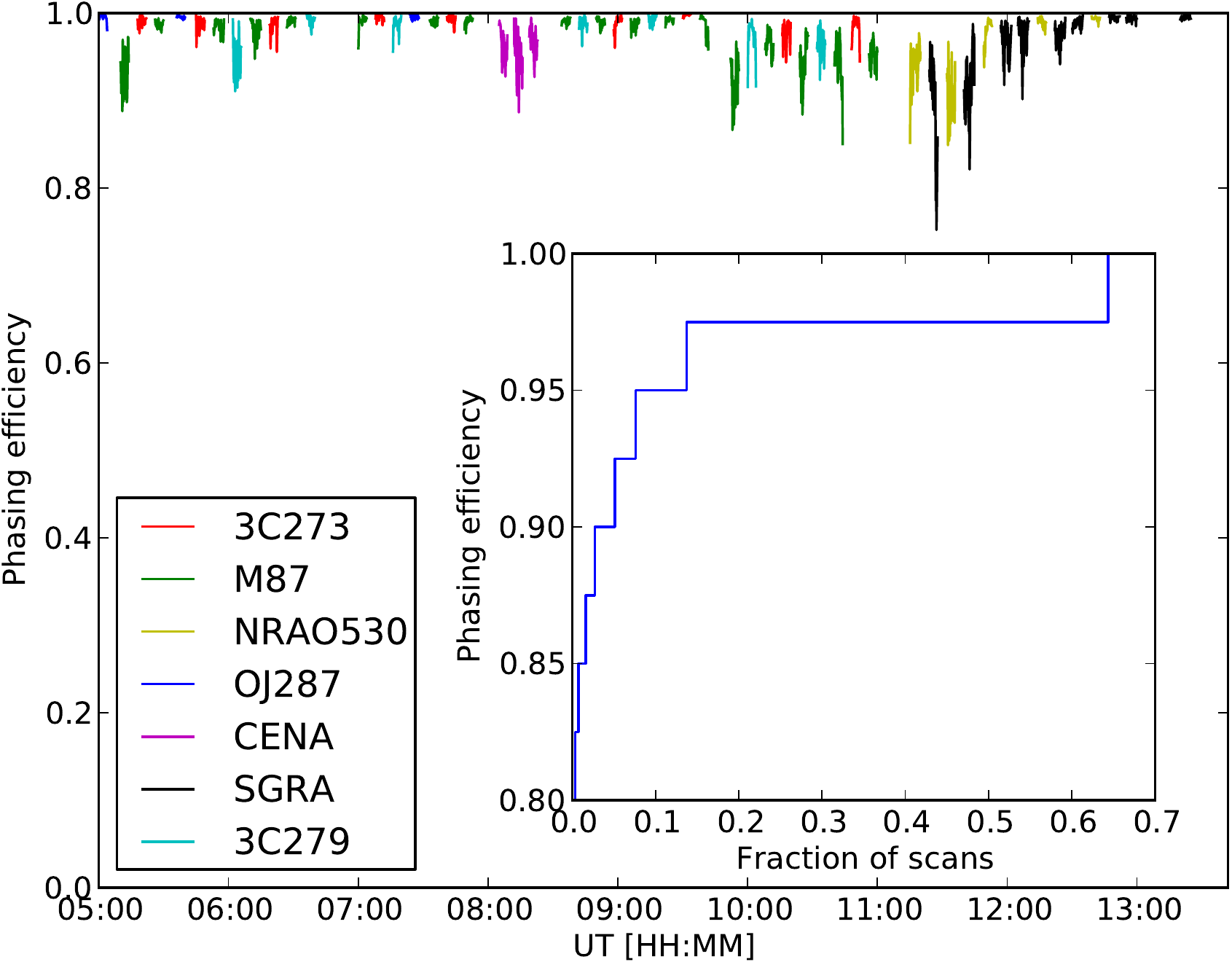}\\
\caption{Phasing efficiency measured on various sources during EHT VLBI on April 4, 2016. The horizontal axis shows time in UT and the vertical axis shows phasing efficiency. The inset histogram shows the distribution of phasing efficiency measured over this period.}
\label{fig:pheff}
\end{center}
\end{figure}

\subsection{SDBE}
Single dish VLBI stations, specifically those used in the EHT in recent years, have a serial data pipeline for 2~GHz bands: digitization, real-time data format to the VLBI standard, encapsulation in the VLBI Data Interchange Format (VDIF) \cite{Whitney09}, and saving data to disk via Mark 6 data recorder \cite{Whitney13}. SWARM distributes the beamformer processing of 2~GHz bands across eight ROACH2 devices. These parallel data streams must be collected and formatted in real-time in order to interface with the Mark 6, in a manner similar to that implemented in the ROACH2 Digital Backend (R2DBE) which is used at other EHT sites \cite{Vertatschitsch_2015}.

\begin{figure}[tbh]
\begin{center}
\includegraphics[width=1.0\textwidth]{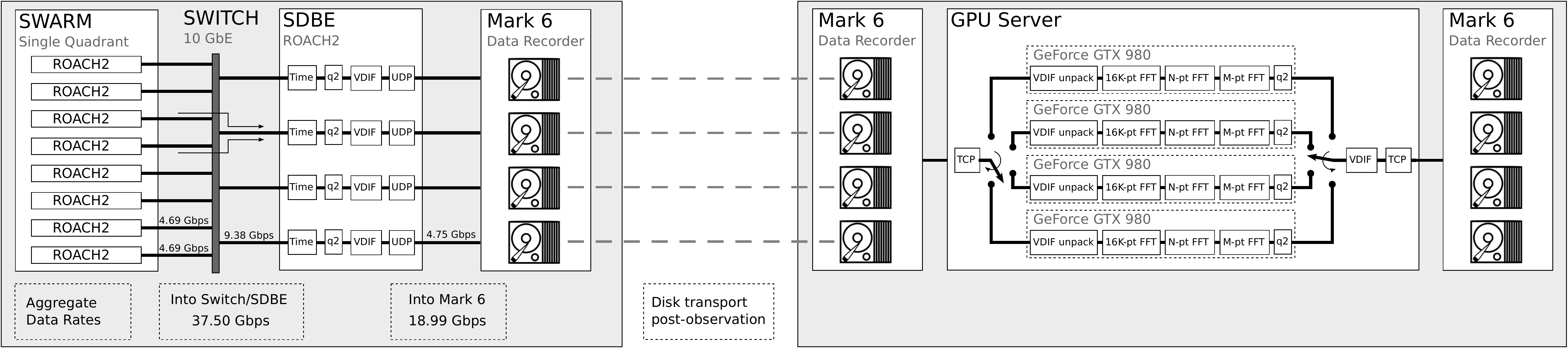}\\
\caption{Block diagram demonstrating the VLBI data pipeline, from SWARM to correlatable data in Mark 6 format. The SDBE is integrated with SWARM and does the real-time processing necessary to interface with the on-site Mark 6 during an observation. After observing the data is preprocessed offline in APHIDS prior to correlation with data from other sites.}
\label{fig:SDBE}
\end{center}
\end{figure}

Utilizing the rapid development platform provided by the ROACH2, we built and tested a real-time system to collect and format ``B-engine'', i.e. beamformer, packets output by SWARM. The data are received on four of the eight 10~GbE ports on the SDBE. The packets are time-stamped, the frequency domain samples are quantized from 4 bits complex to 2 bits complex, the packets are formatted with VDIF headers, and transported over UDP to the Mark 6. Since the B-engine packets are relatively small, several of these packets are bundled into each UDP packet to reduce the interrupt rate on the Mark 6 so as to avoid packet loss. This design uses all eight 10 GbE ports offered on the ROACH2, and all four 10 GbE inputs to the Mark 6. At full speed, the Mark 6 ingests 18.99 Gbps from a single quadrant of SWARM. A block diagram of the SDBE system is shown in Figure~\ref{fig:SDBE}.


\subsection{APHIDS}
\label{sec:aphids}
The underlying data within the packets streamed from the SDBE differs from that typically employed for VLBI and expected at the EHT correlator. Specifically, other EHT sites sample a power-of-two megahertz bandwidth at the Nyquist rate and produce a digital stream of time-domain data. For SWARM the data within the SDBE packets are in the frequency-domain and correspond to a sample rate different from other EHT sites. A certain amount of preprocessing is therefore necessary prior to VLBI correlation with SWARM data, and is performed within APHIDS.

This system reads SDBE data recorded to disk from a Mark 6, converts the data to time-domain at the required sample rate, requantizes to 2 bit, encapsulates in VDIF, and writes to disk on a second Mark 6. The data reformatting implements interpolation and digital filtering using a power-of-two DFT followed by a non-power-of-two inverse DFT, and is GPU accelerated using the CUDA toolset. The filtering discards excess bandwidth resulting from the higher sample rate used in SWARM relative to other sites.

\begin{figure}[tbh]
\begin{center}
\includegraphics[width=0.6\textwidth]{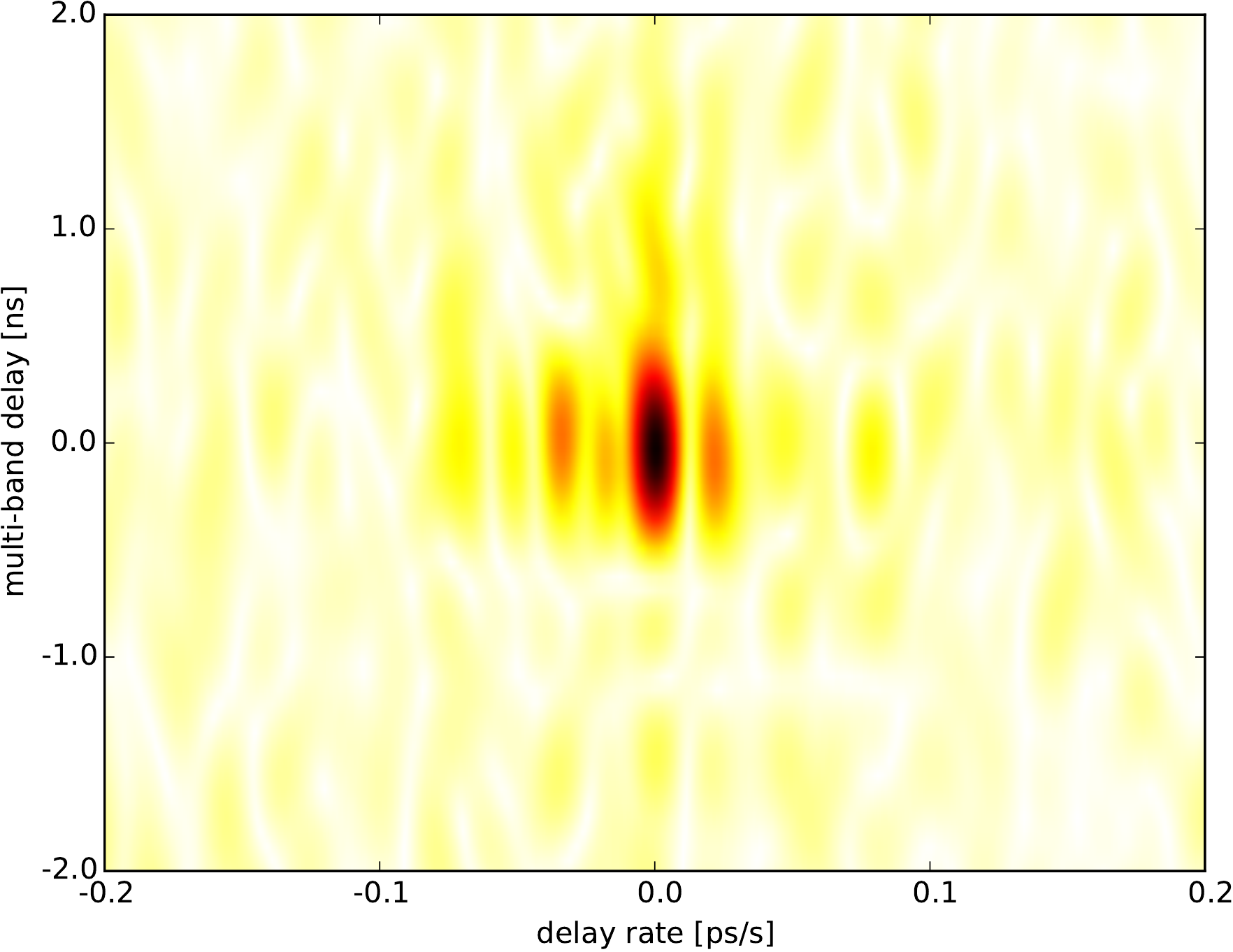}\\
\caption{VLBI fringe detection on the quasar J1512-0905 on a transcontinental baseline between SMA SWARM and the Large Millimeter Telescope (LMT).}
\label{fig:SWARM-LMT}
\end{center}
\end{figure}

Figure~\ref{fig:SWARM-LMT} shows a long baseline fringe detection using SWARM to the Large Millimeter Telescope (LMT) in Mexico, equipped with the ROACH2 DBE. It is typical in VLBI to search for a detection in both delay and delay rate space; the plot shows the correlation coefficient as a function of these variables. Data was taken on 8 April 2016.




\section{Deployment and verification}
 
\begin{figure}[tbh]
\begin{center}
\includegraphics[angle=-0,width=40pc]{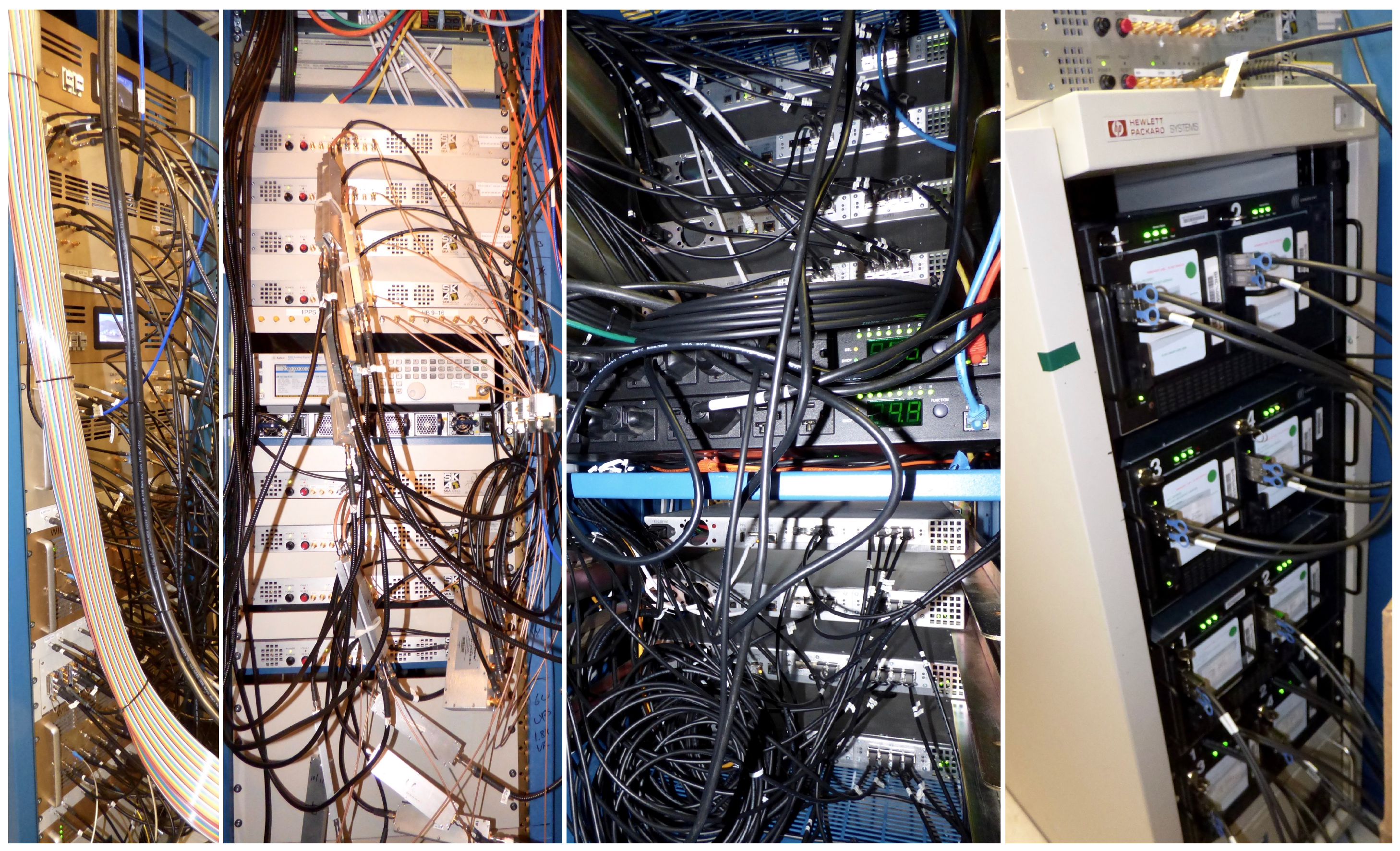}
\caption{Pictures of the SWARM equipment installed on Mauna Kea.  From left to right, the four photos show the BDC which feeds analog baseband signals to SWARM; the front of a single quadrant of SWARM  showing the 8 ROACH2 units cabled with IF, ADC clock, and other control signals entering the front of the ROACH2 chassis; the rear of a SWARM quadrant showing the 10 GbE cables which route the corner turn, visibility and B-engine data; and a rolling rack with a pair of SDBEs and Mark6 data recorders, which record B-engine data from a pair of quadrants. The second installed SWARM quadrant is not shown.  When the SMA ASIC correlator is decommissioned later in 2016, the SWARM equipment in rolling racks (BDC and SDBEs and Mark6 recorders) will be moved to permanent equipment racks.}
\label{fig:swarm}
\end{center}
\end{figure} 

Early prototypes of SWARM were tested in the laboratory starting in 2013 with an antenna simulator, a phase agile four-channel noise generator, with a controllable ratio of correlated to uncorrelated noise in in each channel.  The antenna simulator can be set to Walsh the signals in the characteristic pattern used by the SMA with both 0-180 degree and 90-270 degree cycles.  The simulator could not, however, simulate the geometric delays of a real sky observation.  Also four-antenna versions of SWARM could not correlate the full bandwidth because the cross-multiplies for a single antenna are distributed across all eight ROACH2s  in a full system.  Nonetheless the simulator proved to be invaluable to test basic functionality of SWARM in the laboratory, instead of on the telescope, which requires long distance travel, is less comfortable and efficient due to altitude, and is either constrained in time allocation or risks interfering with SMA observations.

The first eight-ROACH2 quadrant was fielded at the SMA in 2014, running at 54\% of full bandwidth.  Over the next approximately two years, the bandwidth was increased twice, to just over 70\% and then to 90\% in 2015.  Also in late 2015, a second quadrant of SWARM was built and commissioned.  SWARM was first used for EHT VLBI science in March and April 2015, in 70\% bandwidth mode.  It was used again in July 2015 as well as April and June 2016.  VLBI fringe detections were obtained for all these campaigns except June 2016, when a combination of technical problems at the partner EHT site, and bad weather on Mauna Kea, rather than issues with SWARM, obstructed success.

On 11 July 2016, with two quadrants operational, the first full bandwidth bitcode was successfully tested in connected interferometer mode.  On 21 July 2016 two quadrants of SWARM running at full speed were released for science at the SMA. As of 18 October 2016 there are now three quadrants of full speed SWARM in use for science with the ASIC correlator soon-to-be decommissioned. See figure \ref{fig:swarm} for photos of the SWARM equipment installed in the SMA equipment room on Mauna Kea. 

SWARM is always running at full spectral resolution, resulting in data files for a night of observation (assuming the four-quadrant system) of the order of 100~GB in size.  A ``rechunker'' program can quickly reduce the resolution of the SWARM data file, for those who do not need the resolution for their science goals.  The smaller files are more manageable in general, and they load more quickly into the data reduction programs.  Even so,  full resolution SWARM data is archived for every track, which make the archive a more valuable resource when the proprietary period expires and the archive becomes widely available, sometimes for science goals other than that for which the data was originally taken.

\subsection{Line survey demonstration}

\begin{figure}[tbh]
\begin{center}
\includegraphics[width=1.0\textwidth]{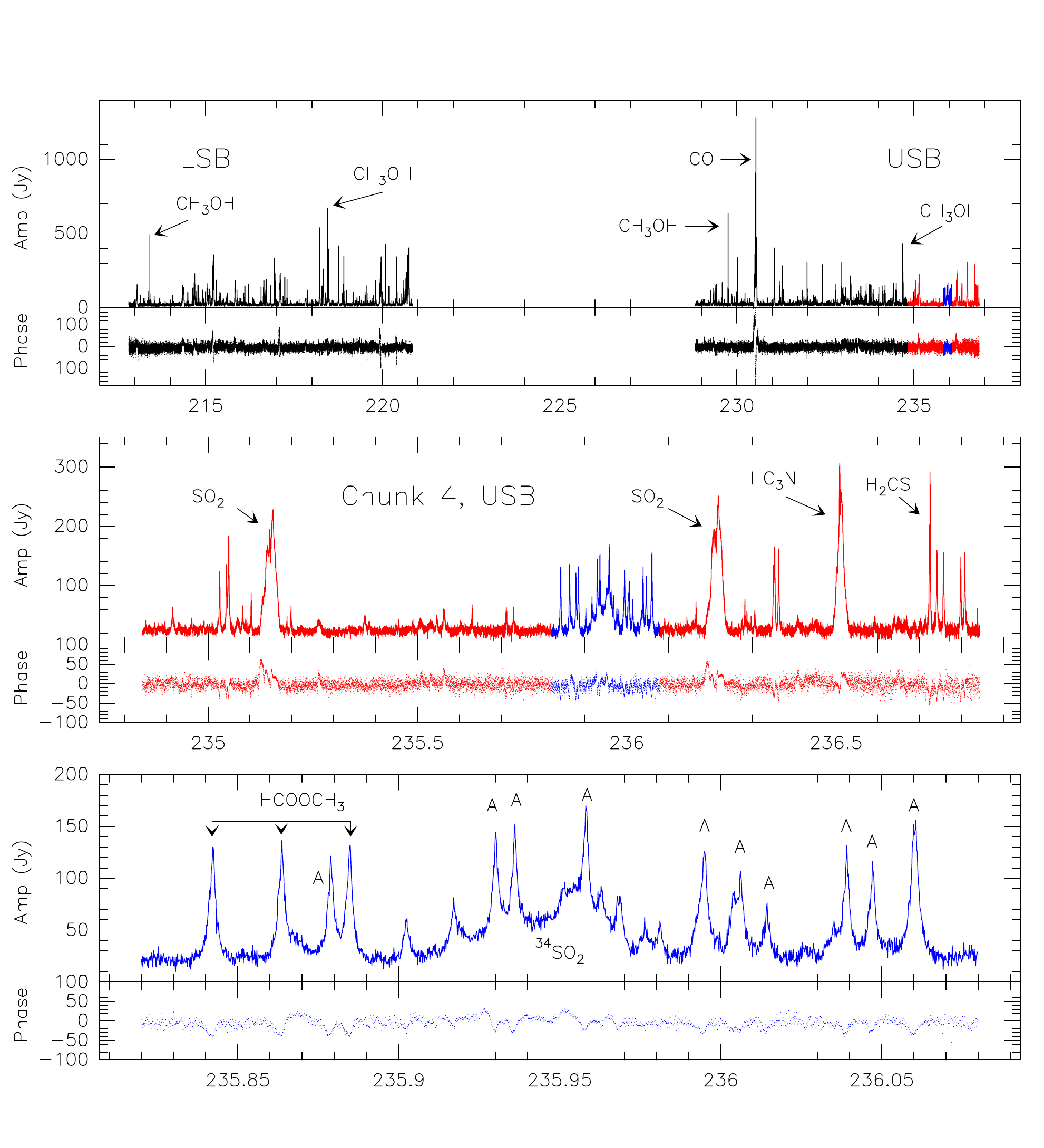}\\
\caption{The SMA with two quadrants of SWARM operational observed the forest of lines in Orion BN/KL on 14 August 2016 between 15:20 and 16:50 UT, early morning in Hawai'i.  The three panels in this presentation zoom in on smaller regions of the spectrum progressively. The top panel shows 16 GHz or 8 GHz in each sideband with the entire band measured instantaneously.    The red section is then blown up in the middle panel, this is a single SWARM 2 GHz chunk in a single sideband.  The blue section is then shown in the bottom panel showing about 260 MHz or about 1.6\% of the observed bandwidth in a two-quadrant SWARM. The lines in the lowest panel marked ``A'' are all transitions of $^{13}CH_3OH$. A single SMA baseline is shown, with one hour of on-source time. The line identifications are from \citet{sutton}.}
\label{fig:swarm_orion}
\end{center}
\end{figure}

The instantaneous wide bandwidth and high spectral resolution of two quadrant SWARM allows for quick and efficient line surveys. To demonstrate this, on 14 August 2016 a verification and demonstration observation, with about an hour on source, was made of the rich forest of strong lines in Orion BN/KL. The opacity was mediocre, $\tau_{225}\sim0.2$, however the atmospheric phase was fairly stable, and the SMA was in the subcompact configuration.  Given the strength of the lines, the conditions were entirely suitable for an observation. Bandpass and flux calibration data was also taken. The calibrated spectrum is show in Figure \ref{fig:swarm_orion}.  

The three panels zoom in on smaller frequency ranges from top to bottom.  The red and blue section in the top panel is a single SWARM 2.0~GHz  usable chunk in one sideband only, and the blue section is a particularly busy and interesting segment of the spectrum, spanning about 260~MHz, and shown in detail in the bottom panel. All of the spectral detail visible in the bottom panel is available across the full spectrum in the top panel, though not well visualized there due to the compressed frequency scale.

When the planned four SWARM quadrants are completed later this year, the 8~GHz gap between lower and upper sidebands apparent in the top panel can be filled (assuming that the two 230~GHz receiver sets are tuned with exactly 8~GHz difference in sky center frequency), and a further 8 GHz contiguous added either below the LSB or above the USB, thereby providing a contiguous 32~GHz instantaneous bandwidth on the sky.  It should also be noted that because SWARM samples  a 2.288~GHz Nyquist band in each ADC channel, and given carefully chosen filters and local oscillators in the block down-converters which condition the IF for SWARM, there are no edge effects every 2~GHz due to bandpass skirts after the guard bands are excised.  In other words the 32~GHz contiguous instantaneous sky band of four-quadrant SWARM when set up in this way has near-optimal SNR anywhere in the band.

\subsection{Quantitative validation}\label{ver:snr}

\begin{figure}[tbh]
\begin{center}
\includegraphics[angle=-0,width=30pc]{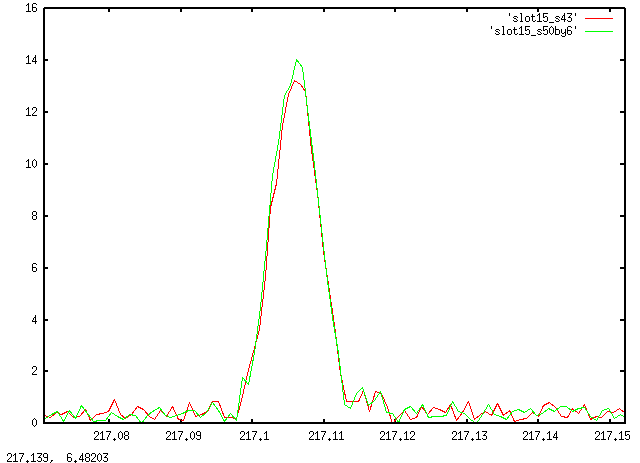}
\caption{ ASIC and reduced resolution SWARM data for one baseline (3-6 of the SiO maser in R-Cas). The ASIC data in chunk s43 is in red, and the SWARM data in chunk s50 is in green. The lines are consistently calibrated and the higher peak in the green line is significant and indicative of the better digital efficiency.}
\label{fig:rcas}
\end{center}
\end{figure} 

To obtain a quantitative measure of SWARM performance, we analyzed observations of the red giant star R Cas taken on 21 July 2016, to see if we get better signal-to-noise with SWARM or ASIC. The SiO(5-4) maser line appeared in ASIC chunk s43, and SWARM chunk s50 (LSB)---for this test the SWARM and ASIC correlators were configured so that both correlators processed the portion of the IF which contained the spectral line. There was no detectable continuum, and no other lines, so this observation lends itself well to a comparison of SNR in the two systems---SNR in this context is defined as the ratio of line area to the Root-Mean-Square (RMS) of the system noise in the line free region of the spectrum. See Figure \ref{fig:rcas} for the SiO maser line as seen by SWARM and the SMA ASIC correlator.

A short form description of the analysis is given in the following steps.  Standard SMA data calibration (steps 1 to 3) used the SMA data reduction package, MIR. Python code was used to complete the analysis and estimate the SNR (steps 4 to 9).

\begin{arabiclist}
\item $T_{\mathrm{sys}}$ calibration was applied to the data, using SMA's logged Y-factor measurements of system temperature.   This converted the raw crosscorrelation coefficient to an approximate Jansky unit scale.
\item Bandpass calibration was completed in both data sets using data taken on the quasar 3c454.3 
\item The s43 (ASIC) and s50 (SWARM) R Cas amplitude data was gain calibrated  using  MWC349 as a calibration source
\item The SWARM data was vector averaged in sets of 6 channels to approximately match the ASIC resolution
\item RMS values of the amplitude of these spectra were calculated in the frequency range corresponding to the usable 82~MHz bandwidth of s43 (excluding the region of line emission) with s50 trimmed to the same 82 MHz to get the ASIC RMS.
\item The average value of the amplitude in the line was calculated for each spectrum
\item The SNR was calculated by dividing the average line-area amplitude by the RMS.
\item The ratio of the s50 SNR to the s43 SNR was calculated for each baseline.
\item The average of all 28 SNR ratios from step 8 then showed a ratio of 1.11 with an error of $\pm 0.03$.
\end{arabiclist}
A standard 2-bit correlator has a digital efficiency of 0.88 compared to a continuous, i.e. analog, correlator. When the lower products are dropped, as is the case for the ASIC, the efficiency drops to 0.87 relative to an analog correlator. SWARM, which is actually a 4-bit correlator although it samples at 8-bits, should have a digital efficiency of 0.99 compared to an analog correlator and thus should see a $\sim$12\% improvement in SNR over the ASIC correlator.
This analysis shows the measured SNR for SWARM is $11\% \pm 3\%$ higher than for the ASIC. This non-trivial improvement allows SWARM to achieve ASIC's SNR with correspondingly less telescope time.

\subsection{Sample science data}

\begin{figure}[tbh]
\begin{center}
\includegraphics[angle=-0,width=40pc]{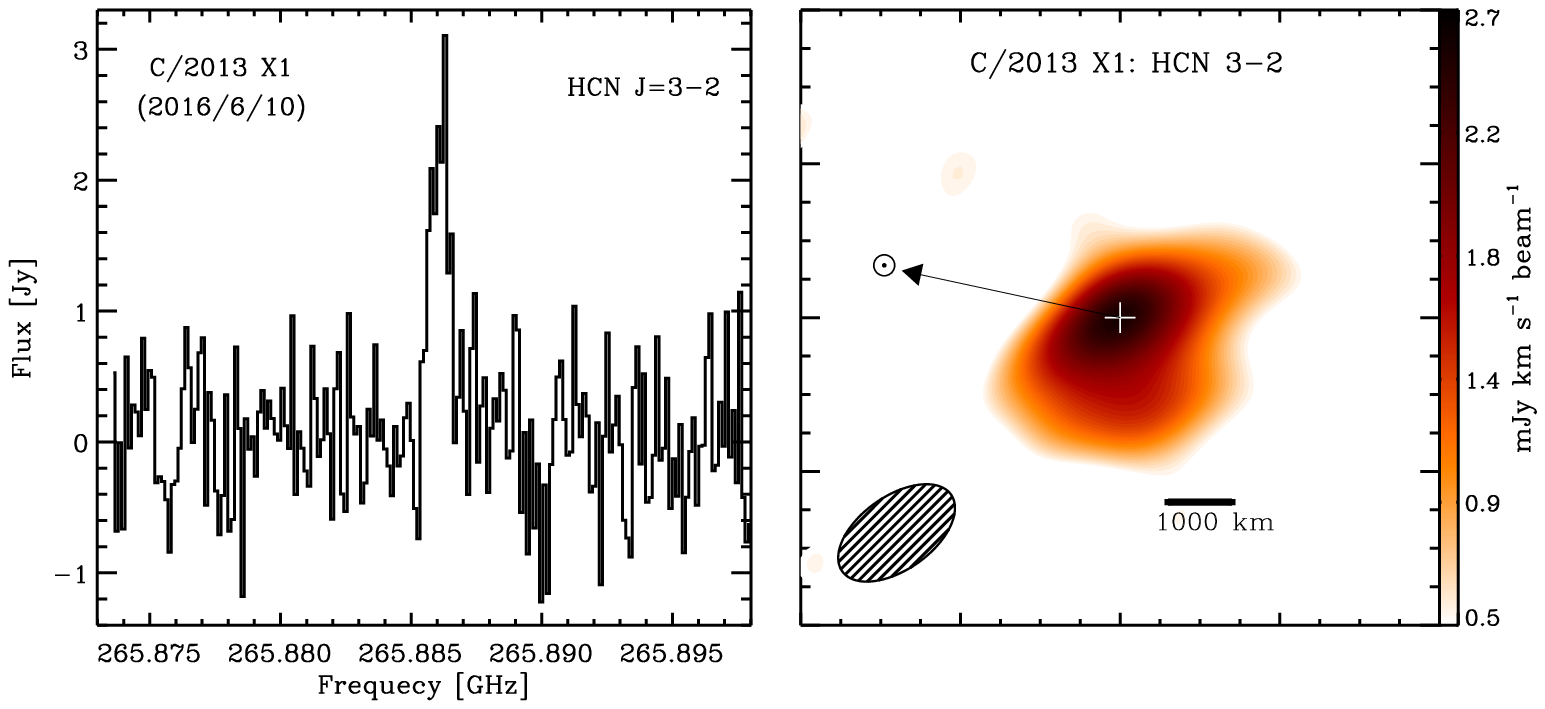}
\caption{ HCN J=3 $\mapsto$ 2 spectrum (left) and image (right) toward comet C/2013 X1 (PanSTARRS). The observations were taken on 10 June 2016  using the dual receiver SWARM mode with a spectral resolution of 127 kHz. This narrow line (left) is resolved with SWARM's high uniform spectral  resolution. The integrated intensity image shows the HCN emission peaked around the nucleus position but with clear extension toward the anti-solar direction. The cross marks the position of the comet's nucleus and the arrow shows the direction of the Sun. North is the positive Y-axis  and East is the positive X axis. The data were reduced and kindly shared by Chunhua Qi }
\label{fig:comet}
\end{center}
\end{figure} 

SWARM is routinely used for science.  Figure \ref{fig:comet} shows a narrow (about 1~MHz) spectral line and line image of a HCN transition in a comet, observed by Smithsonian Scientist Chunhua Qi.  The line takes up about seven 130~kHz SWARM bins, it was taken when SWARM was one step away from running at full bandwidth.

\section{Conclusions} \label{conclusions}

We have built and commissioned three quadrants of the SWARM system processing 24~GHz of the eventual 32~GHz bandwidth goal. The three quadrants fully validate the SWARM design since the quadrants are essentially replicas of one another. The two SWARM instruments, the connected correlator and the phased array, have been successfully deployed for routine science, and represent the future of digital signal processing at the SMA. The older ASIC correlator will be retired in 2016, saving an order of magnitude in power used for DSP at SMA, and freeing up space in the SMA correlator room for future instrument build outs. Engineering decisions made early in the design process of SWARM that have been validated include:

\begin{itemlist}
\item The use of quad-core ADCs in a broadband application which was viewed as a technical risk in the CASPER community.  The foundational work of \citet{patel} on mitigation of distortion through quad core alignment has allowed us to show that such devices can yield science quality wideband astronomical data.  
\item The choice to build an FX correlator with high resolution spectral decomposition computed in one DSP stage, since cascading coarse and fine PFBs would cause edge effects requiring overlapping coarse channels to mitigate, creating a need for still more computation, more FGPA hardware, and complex interconnect. Early utilization estimates showed that two 32~kilopoint  PFBs would fit on a single Xilinx FPGA, with X-engines co-located, along with delay and phase alignment, networking, packetizing, and transpose and buffer resources, as long as the demultiplex factor was limited to 16.   
\item The choice of a demultiplex factor of 16 along with the chunk bandwidth of 2.3~GHz which necessitated an FPGA clock rate of 286~MHz and very high utilization of the various FPGA resources. Meeting timing was indeed a greater challenge than anticipated but was ultimately achieved in July 2016.
\item The election to use open-source CASPER technology, including the ROACH2 and 5~GSps ADCs. The SMA internal design efforts were limited to system design, infrastructure, and the very complex, highly utilized and high performance, FPGA bit code.  We did not, however, have to develop and debug DSP hardware, which would have resulted in longer ``time to science'' for SWARM.
\end{itemlist}
All the originally targeted goals set at project inception were achieved.  SWARM is impressively full featured, compact, and economical in its power consumption, and while these desirable characteristics are in part a consequence of Moore's Law, some were met through persistent pursuit of an elegant, highly utilized, and challenging high speed FPGA design.  

Though this is not the first CASPER packetized correlator, it is to our knowledge the widest bandwidth CASPER correlator deployed as an open facility instrument, therefore further validating CASPER approaches such as the use of of packet-switched Ethernet switch based corner turners, and the benefits of open source sharing of technology within the Astronomical community.



\section*{Acknowledgments}
The Submillimeter Array is a joint project between the Smithsonian Astrophysical Observatory and the Academia Sinica Institute of Astronomy and Astrophysics. We are grateful for the hard work and support of numerous SMA staff, who, collectively, made SWARM possible. Development of the VLBI features of SWARM were funded with SAO Internal Research \& Development funding, the NSF, and the Gordon and Betty Moore Foundation under GBMF3561. We received generous donations of FPGA chips from Xilinx, Inc, under the Xilinx University Program, also supporting EHT VLBI SWARM features. We acknowledge the EHT for providing the SWARM EHT fringe verification data, and Chunhua Qi for the HCN line and image plot in comet C/2013 X1 (PanSTARRS).  SWARM has benefited from technology shared under open source license by CASPER.  This research has made use of NASA's Astrophysics Data System. We acknowledge the significance that Mauna Kea has for the indigenous Hawaiian people, and are privileged to be able locate SWARM at its summit.


\begin{thebibliography}{10}

\bibitem[Escoffier {\it et al.} (2007)]{escoffier} Escoffier, R. P., Comoretto, G., Webber, J. C., Baudry, A., Broadwell, C. M., Greenberg, J. H., Treacy, R. R., Cais, P., Quertier, B., Camino, P., Bos, A., and Gunst, A. W. {\bf A\&A } 462, 801-810 (2007).

\bibitem[Ho {\it et al.} (2004)]{smaho} Ho, P. T. P., Moran, J. M., and Lo, K. Y. {\bf ApJ}, 616, L1 (2004).

\bibitem[Jiang {\it et al.} (2014)]{jiangadc} Jiang, H., Liu, H., Guzzino, K., Kubo, D., Li, C.-T., Chang, R., and Chen, M.-T., {\bf PASP}, 126:761-768 (2014).    

\bibitem[Johnson {\it et al.} (2015)]{Johnson_2015} Johnson, M. D., Fish, V. L., Doeleman, S. S., {\it et al.}, {\bf Science}, 350, 1242 (2015).
  
\bibitem[Parsons {\it et al.} (2008)]{parsonsa} Parsons, A., Backer, D., Chen, H., Droz, P., Filiba, T., MacMahon, D., Manley, J., McMahon, P., Parsa, A., Siemion, A., Werthimer, D., Wright, M., {\bf PASP}, 120, 873, 1207-1221 (2008).

\bibitem[Parsons {\it et al.} (2005)]{parsonsb} Parsons, A., {\it et al.}, ``A New Approach to Radio Astronomy Signal Processing'', {\bf URSI GA}, (2005).

\bibitem[Patel {\it et al.} (2014)]{patel} Patel, N. A., Wilson., R. W., Primiani, R. A., Weintroub, J., Test, J. and Young, K. H., {\bf Journal of Astronomical Instrumentation}, Vol. 3, No. 1 (2014).


\bibitem[Sutton {\it et al.} (1985)]{sutton} Sutton, E. C., Blake, G. A., Masson, C. R., and Phillips, T. G., {\bf ApJS}, Vol. 58, p341-378 (1985).


\bibitem[Vertatschitsch {\it et al.} (2015)]{Vertatschitsch_2015} Vertatschitsch, L., Primiani, R. A., Young, A., {\it et al.}, {\bf PASP}, 127, 1226 (2015).

\bibitem[Whitney {\it et al.}(2009)]{Whitney09} Whitney, A. R., Kettenis, M., Phillips, C. and Sekido, M. [2009] ``VLBI Data Interchange Format (VDIF),'' {\bf Proceedings of the 8th International e-VLBI Workshop}, Vol. 42 (2009).

\bibitem[Whitney {\it et al.}(2013)]{Whitney13} Whitney, A. R., Beaudoin, C. J., Cappallo, R. J., Corey, B. E., Crew, G. B., Doeleman S. S., Lapsley, D. E., {\it et al.}, ``Demonstration of a 16 Gbps Station-1 Broadband-RF VLBI System,'' {\bf PASP}, Vol. 125, No. 924, (2013).

\bibitem[Young {\it et al.}(2016)]{Young16} Young, A., Primiani, R. A., Weintroub, J., Moran, J. M., Young, K. H., Blackburn, L., Johnson, M. D., and Wilson, R. W., ``Performance Assessment of an Adaptive Beamformer for the Submillimeter Array.''  {\bf  IEEE International Symposium on Phased Array Systems \& Technology}, in press, (2016).

\end{thebibliography}
\end{document}